\documentclass[prl,twocolumn,amsmath,amssymb,amsfonts,superscriptaddress,floatfix,showpacs,aps,longbibliography]{revtex4-1}
\newcommand{\beq}{\begin{equation}}
\newcommand{\eeq}{\end{equation}}
\usepackage{graphics}
\usepackage{amsmath}
\usepackage{amsfonts}
\usepackage{amssymb}
\usepackage{graphicx}
\usepackage{color}
\usepackage[normalem]{ulem}
\usepackage{notes2bib}

\usepackage{hyperref} 
\usepackage[group-separator={,}]{siunitx} 
\usepackage[english, capitalise]{cleveref} 

\begin{document}

\title{X-ray View of Light-Induced Spin Reorientation in TmFeO$_{3}$: Direct Observation of a 90$^\circ$ Néel Vector Rotation}

\author{Somnath Jana}
\email[Email: ] {somnath.jana@mbi-berlin.de/sj.phys@gmail.com}
 \affiliation{Max Born Institute for Nonlinear Optics and Short Pulse Spectroscopy, Max-Born-Str. 2a, 12489 Berlin, Germany}
\author{Ronny Knut}
 \affiliation{Department of Physics and Astronomy, Uppsala University, Box 516, 75120 Uppsala, Sweden}
\author{Dima Afanasiev}
 \affiliation{Institute for Molecules and Materials, Radboud University, Nijmegen, The Netherlands}
\author{Niko Pontius}
 \affiliation{Helmholtz-Zentrum Berlin für Materialien und Energie GmbH, Albert Einstein Straße 15, 12489 Berlin, Germany}
\author{Christian Schüßler-Langeheine}
 \affiliation{Helmholtz-Zentrum Berlin für Materialien und Energie GmbH, Albert Einstein Straße 15, 12489 Berlin, Germany}
\author{Christian Tzschaschel}
 \affiliation{Max Born Institute for Nonlinear Optics and Short Pulse Spectroscopy, Max-Born-Str. 2a, 12489 Berlin, Germany}
 \affiliation{Department of Physics, University of Zurich, Zurich, Switzerland}
\author{Daniel Schick}
 \affiliation{Max Born Institute for Nonlinear Optics and Short Pulse Spectroscopy, Max-Born-Str. 2a, 12489 Berlin, Germany}
 \author{Alexey V. Kimel}
  \affiliation{Institute for Molecules and Materials, Radboud University, Nijmegen, The Netherlands}
\author{Olof Karis}
 \affiliation{Department of Physics and Astronomy, Uppsala University, Box 516, 75120 Uppsala, Sweden}
 \author{Clemens von Korff Schmising}
 \affiliation{Max Born Institute for Nonlinear Optics and Short Pulse Spectroscopy, Max-Born-Str. 2a, 12489 Berlin, Germany}
\author{Stefan Eisebitt}
 \affiliation{Max Born Institute for Nonlinear Optics and Short Pulse Spectroscopy, Max-Born-Str. 2a, 12489 Berlin, Germany}
 \affiliation{Technische Universität Berlin, Straße des 17. Juni 135, 10623 Berlin}

\begin{abstract}
Using time-resolved X-ray magnetic linear dichroism in reflection, we provide a direct probe of the Néel vector dynamics in TmFeO$_3$ on a ultrafast timescale. 
Our measurements reveal that, following optical excitation, the Néel vector undergoes a spin reorientation transition primarily within the a–c plane, completing a full 90° rotation within approximately 20 ps. This study highlights the ability to probe dynamics of antiferromagnets at its intrinsic timescale in reflection geometry, paving the way for investigations of a wide range of antiferromagnets grown on application relevant substrates. 
\end{abstract}

\maketitle

\section{Introduction}
Antiferromagnets (AFMs) are increasingly recognized as promising candidates for next-generation spintronics and magnonics applications, offering key advantages over conventional ferromagnetic (FM) materials~\cite{Baltz2018, MacDonald2011, Gomonay2014}. One of their most compelling feature is their ultrafast spin dynamics, which reside in the terahertz (THz) frequency range~\cite{Keffer1952}. This makes AFMs highly attractive for high-speed information processing and data storage.  However, the absence of net magnetization of AFMs significantly hinders their integration into modern technologies, which places the manipulation and reading of the AFM states at the center of ongoing research~\cite{Wadley2016, Baldrati2019, Selzer2022, PhysRevLett.113.157201, Jong2011, Kubacka_Staub2014, Mikhaylovskiy2015, Baierl2016, Zhang2016, ThielemannKuhn_2017, Neamec2018, Afanasiev2021,  Fitzky2021}. 
In the absence of net magnetization, the Néel vector is introduced as a new order parameter, enabling the definition of memory states. Several recent experimental studies on Néel vector switching have relied on current pulses~\cite{Wadley2016, Baldrati2019, Selzer2022, PhysRevLett.113.157201}, which operate on much slower time scales than the intrinsic AFM dynamics. Common probing techniques, such as anisotropic magnetoresistance and spin Hall magnetoresistance read-outs, are also limited to much slower time scales and may be affected by thermal artifacts~\cite{Chiang2019, Churikova2020,Cheng2020_alphaFe2O3, Zhang2019_Fe3O4_Pt}. While ultrashort optical pulses have been used to investigate AFM dynamics~\cite{Kimel2004, Jong2011, Kalashnikova2007, Tzschaschel2017}, sub-picosecond processes remain largely unexplored, possibly due to the state-filling effect influencing the optical response at these time scales~\cite{Koopmans2000_MagnetismOrOptics}. 

A direct and artifact-free approach to studying AFM dynamics at intrinsic time scales involves combining ultrashort optical or THz excitation~\cite{Kampfrath2011_THz, Kubacka_Staub2014, Baierl2016}  with femtosecond X-ray probing. X-ray Magnetic Linear Dichroism (XMLD) offers element-specific sensitivity through core-level transitions, allowing for tracking of local magnetic moments at relevant absorption edges. It arises from an anisotropic charge distribution, induced by exchange and spin-orbit interactions, which is proportional to the square of the sublattice magnetization and therefore also exists for AFMs. While XMLD has become a powerful technique for studying AFMs in equilibrium~\cite{kuiper1993,kortright2000,Oppeneer2003_BuriedAFM, kunes2003}, its application to ultrafast, out-of-equilibrium states remains largely unexplored~\cite{Harris2023}. 

\begin{figure}[]
\includegraphics[width=0.99\columnwidth]{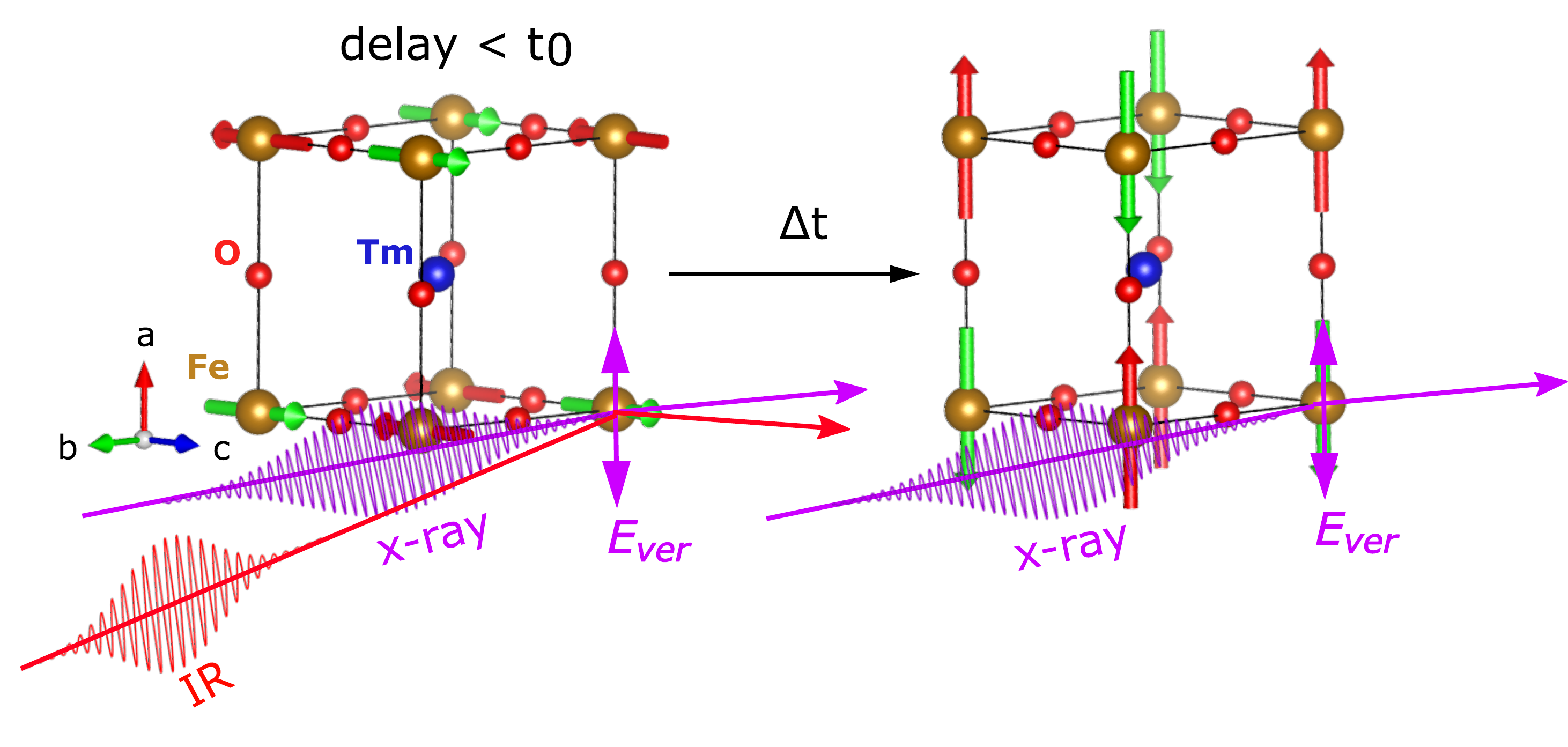}
\caption{(Color Online) Approximate crystal structure and spin orientation of TmFeO$_3$ below (left) and above (right) the SRT temperature. The schematic demonstrate the measurement geometry used in this experiment. The red and violet colors are used to denote the NIR and x-ray pulse and beam directions. The double-headed violet arrows indicate the vertical mode of the x-ray polarization.  \label{Fig1}}
\end{figure}

Among various AFMs, REO are known for their spin reorientation phase transitions (SRT) \cite{White1969, Leake1968, Pinto1971, Shapiro2014, Staub2017}. The SRT in TmFeO$_3$ is characterized by two second-order phase transitions occurring at 80 K and 90 K. The high-spin state Fe$^{3+}$ sublattice orders in a G-type antiferromagnetic structure below the Néel temperature of $\sim$ 630 K, with the Néel vector initially aligned along the a-axis (cf. Fig. \ref{Fig1}, right). As the system is cooled, the spin reorientation begins near 90 K and completes around 80 K, with the Néel vector aligning along the c-axis (cf. Fig. \ref{Fig1}, left). 
Microscopically, the SRT is driven by changes of the magnetocrystalline anisotropy, allowing a laser-driven manipulation of the Néel vector via spin-lattice coupling, as was shown by Kimel et al.~\cite{Kimel2004}.

To directly probe the Néel vector dynamics during laser-driven spin reorientation in TmFeO$_3$, we performed time-resolved XMLD measurements in reflectivity using ultrashort ($\Delta t \sim 100$\,fs) x-ray pulses. Measuring XMLD in reflection can yield significantly stronger contrast~\cite{Oppeneer2003_BuriedAFM, Tesch_2014}, compared to transmission. Moreover, reflection-based measurements eliminate the requirement for x-ray-transparent substrates, broadening the range of compatible material systems and simplifying sample preparation. Our results reveal a well-defined rotation of the Néel vector within the ac-plane, completing its transition from the c-axis to the a-axis in about 20\,ps.  This observation provides a direct, time-resolved view of the spin reorientation process, free from artifacts associated with optical or electrical detection methods. The study demonstrates the ability of XMLD for probing ultrafast spin dynamics in antiferromagnets, offering a valuable tool for advancing AFM-based spintronic technologies.

\section{Experimental methods}
\begin{figure}[]
\includegraphics[width=0.99\columnwidth]
{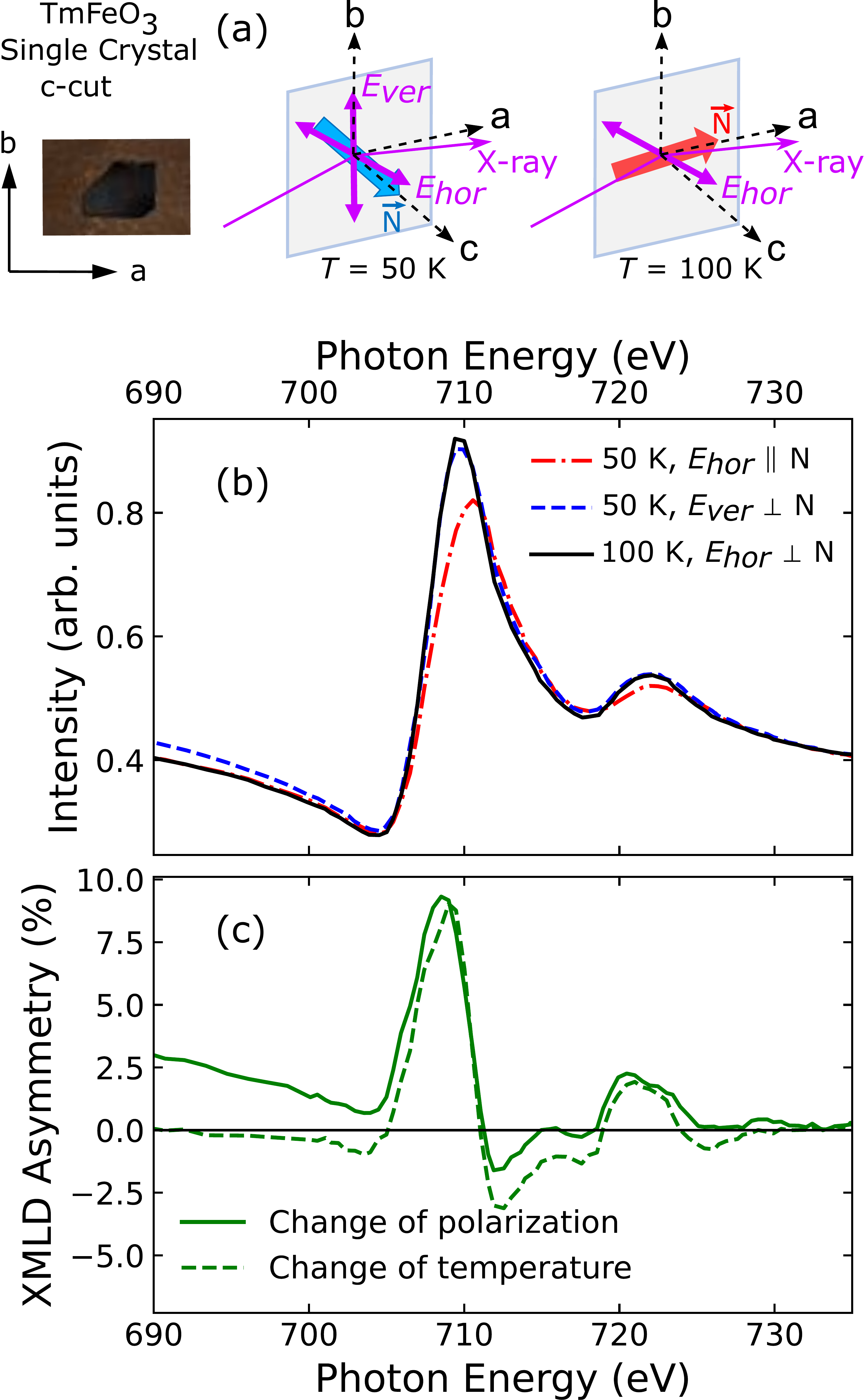}
\caption{(Colour Online) (a) Image of the TmFeO$_3$ single crystal used in the experiment, along with the measurement geometries for static XMLD measurements. The plane of incidence is spanned by the a-c plane, while x-ray polarization can be changed between $E_{hor}$ (forming 5$^{\circ}$ angle with c-axis in the a-c plane) and $E_{ver}$ (normal to a-c plane) directions. (b) X-ray reflectivity spectra measured at 50 K for both horizontal and vertical polarizations, and at 100 K for horizontal polarization. (c) XMLD asymmetry obtained using two approaches: varying x-ray polarization while keeping the spin orientation fixed along c-axis (solid line) and varying spin orientation by changing temperature while keeping the x-ray polarization fixed at horizontal (dashed line). \label{Fig2}}
\end{figure}

A TmFeO${_3}$ single crystal was prepared using the floating zone technique and cut along to the c-crystal axis in the form of a 60\,$\mu$\text{m} thick plate of size $\sim$ 4$\times$3 mm$^2$ (cf. Fig. \ref{Fig2}(a)), which was then optically polished. 
NIR pump and x-ray probe experiments were conducted at the Femtoslicing facility (beamline UE56/1-ZPM and DynaMaX endstation) at BESSY II, Helmholtz-Zentrum Berlin. Transient XMLD measurements were performed in reflection utilizing both the normal and slicing modes of the beamline. The temporal resolution in the normal mode is $\sim$ 70 ps, determined by the approximate temporal width of the electron bunch, while the slicing mode generates ultrashort X-ray pulses of $\sim$ 100 fs through the interaction between an electron bunch and an intense near-infrared (NIR, center wavelength of 800 nm) pulse~\cite{Holldack2014}. The probing frequency is 6 kHz, while sample excitation is performed at 3 kHz, allowing for alternating measurement of excited (pumped) and unexcited (unpumped) signals using an avalanche photo-diode. This approach effectively eliminates noise and drifts occurring on timescales longer than 3 kHz, enabling accurate measurements despite the low flux of about 10$^6$ photons/second on the sample in the slicing mode. The X-ray energy resolution was $E/{\Delta E} \approx 500$.

Figure \ref{Fig2}(a) displays an image of TmFeO$_3$, on which the measurements were performed. 
The experiment was conducted in reflection geometry at a grazing incidence angle of 5$^{\circ}$. Consequently, the c-axis forms an 85$^{\circ}$ angle with the incident x-ray beam (cf. Fig. \ref{Fig2}(a)). The orientation of the a- and b-axes relative to the x-ray polarization was adjusted to probe the Néel vector. The sample was cooled to below the spin reorientation temperature using liquid helium and the temperature was monitored using a Si-diode sensor mounted on the Cu holder near the sample. The XMLD asymmetry is calculated by taking the normalized difference between the reflectivities measured with linearly polarized X-rays aligned parallel and perpendicular to the spin axis.

\section{Results and discussions}
First, static measurements were carried out using two different approaches to determine the XMLD asymmetry between the two spin orientations. This asymmetry serves as a reference for evaluating NIR pulse-induced changes in the transient spectra and for quantifying the angle of the spin-reorientation. In the first approach, the sample was cooled to \SI{50}{K}, and spectra were recorded across the Fe L$_{2,3}$ edges for horizontal ($E_\mathrm{hor}$) and vertical ($E_\mathrm{ver}$) polarizations with respect to the plane of incidence (cf. Fig. \ref{Fig2}(a). Since the sample temperature remained below the SRT temperature, the Néel vector remained aligned along the c-axis. Changing the polarization leads to variations in reflected intensity at the absorption edges due to XMLD.  
We correct the measured spectra for the experimentally determined difference in beamline transmission for the two X-ray polarizations and show the results in Fig. \ref{Fig2}(b). 
The corresponding asymmetry is shown in Fig. \ref{Fig2}(c). A clear XMLD signal is observed at both edges, with a maximum asymmetry of approximately 9\% at a photon energy $\sim$ 1\,eV below the L$_3$ edge maximum. 

To further confirm and quantify the XMLD asymmetry, a second approach is taken in which the X-ray polarization is fixed horizontally and spectra are compared at temperatures below and above the SRT temperature. 
For this purpose, an additional spectrum was recorded at \SI{100}{K} for horizontal polarization, which is presented in Fig. \ref{Fig2}(b) (solid line). 
At 100 K (above SRT), the Néel vector is expected to align along the a-axis, forming a 85$^\circ$ angle with the x-ray polarization, $E_\mathrm{hor}$ (cf. Fig. \ref{Fig2}(a)). 
Consequently, the spectrum closely resembles the one measured at \SI{50}{K} with vertical x-ray polarization (dashed line), where the Néel vector, aligned along the c-axis, forming a 90$^\circ$ angle with $E_\mathrm{ver}$. 
As we observe that over time a thin layer of condensation forms on the cold sample, we correct for the associated change of reflectivity by normalizing the spectrum measured at \SI{100}{K} to the pre- and post-edge reflectivities of the spectrum obtained at \SI{50}{K}.
The XMLD asymmetry obtained in this approach is presented in Fig. \ref{Fig2}(c) (dashed line). 
Clearly, both approaches lead to very similar XMLD asymmetry spectra, with a maximum of about 8.5\% at $\sim$ 708.5\,eV. 

\begin{figure}[]
\includegraphics[width=0.99\columnwidth]
{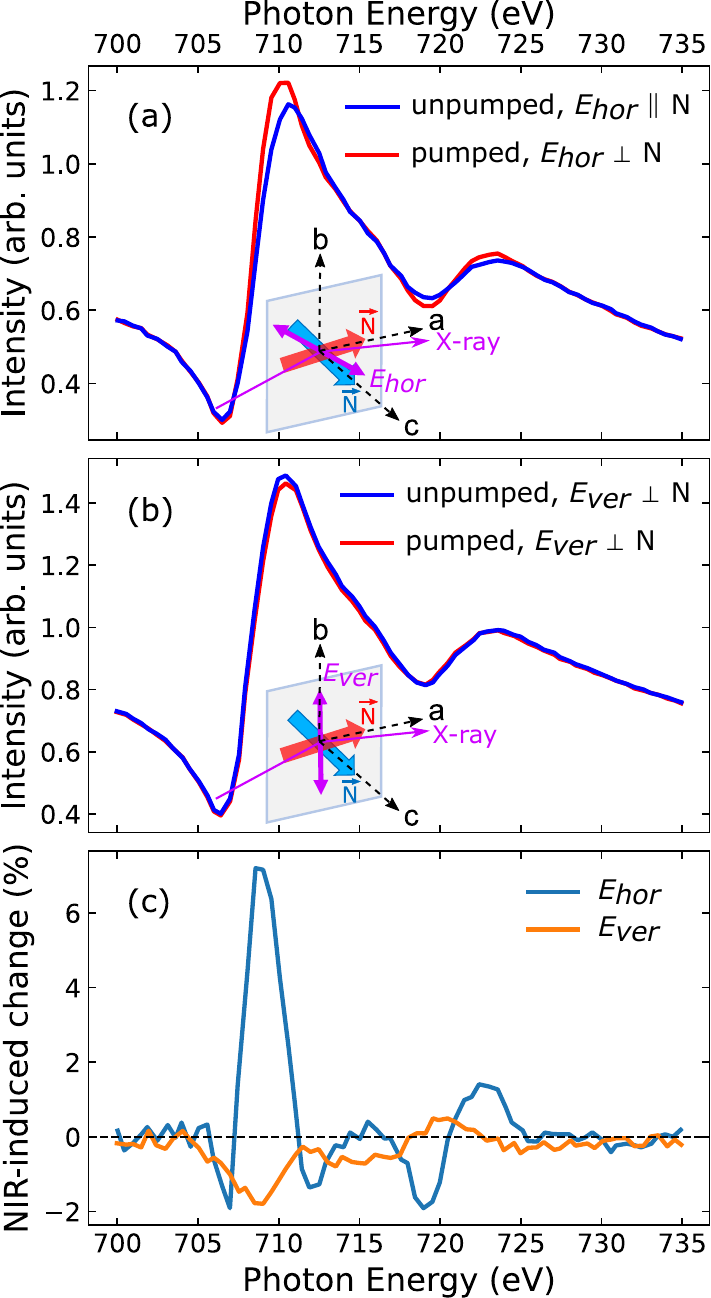}
\caption{(Colour Online) (a) X-ray reflectivity spectra for the excited (pumped) and ground (unpumped) states measured with horizontal polarization. (b) Same as (a), but for vertical polarization. (c) NIR pump-induced change derived from the spectra in (a) (blue) and (b) (orange). \label{Fig3}}
\end{figure}
In time-resolved measurements, the sample is excited by NIR laser pulses with an incident fluence of 45 mJ/cm$^2$.
In this wavelength range the absorption is dominated by the electronic excitations of the iron ions \cite{usachev2005}.  
The NIR-induced changes in the reflectivity signal are first investigated in the the normal mode with a temporal resolution of 70 ps. 
The initial sample temperature is kept constant at $\sim$~32 K, well below the  SRT temperature.  
The time-delay is set to \SI{170}{ps} after laser excitation, a time for which we can assume the sample to be in thermal equilibrium at an elevated temperature. 
To derive the changes in XMLD asymmetry, measurements were conducted for both horizontal and vertical X-ray polarizations.
X-ray reflectivity spectra were measured across the Fe $L_{2,3}$ edges for both the pumped and unpumped states of the sample and are presented in Figs. \ref{Fig3}(a) and (b) for the horizontal and vertical x-ray polarizations, respectively.  Since the pumped and unpumped spectra were recorded using consecutive X-ray pulses from the 6 kHz repetition rate source, no additional normalization was required to compare the spectra. Furthermore, we observed no intensity changes due to condensation in the presence of the NIR pulse. 

Figure \ref{Fig3}(c) shows the normalized difference between the pumped and unpumped spectra displayed in Figs. \ref{Fig3}(a) and (b), highlighting the pump-induced effect. This spectral change closely resembles the characteristic XMLD features of Fe$^{3+}$ in an octahedral crystal field~\cite{Stoehr_Siegmann, Staub2017}, as well as the static XMLD asymmetry shown in Fig. \ref{Fig2}(c), indicating a laser-induced reorientation of the Néel vector. For horizontal x-ray polarization, a maximum change of approximately 7\% is observed at the Fe L$_3$ edge  corresponding to about 85\% of the static XMLD asymmetry measured between the two ground states below 80 K and above 90 K (cf. Fig. \ref{Fig2}(c)). This result provide direct evidence of optical laser-induced switching of the Néel vector from the c-axis to the a-axis. For vertical x-ray polarization, the XMLD asymmetry is significantly weaker and exhibits an opposite sign. In this case, the angle between the Néel vector and the x-ray polarization remains unchanged, regardless of whether the spins are aligned along the c-axis or a-axis. The small negative XMLD signal observed may be attributed  
to a thermally induced spin-excitations leading to a finite projection of the spin-axis along $E_\mathrm{ver}$.

\begin{figure}[]
\includegraphics[width=0.99\columnwidth]{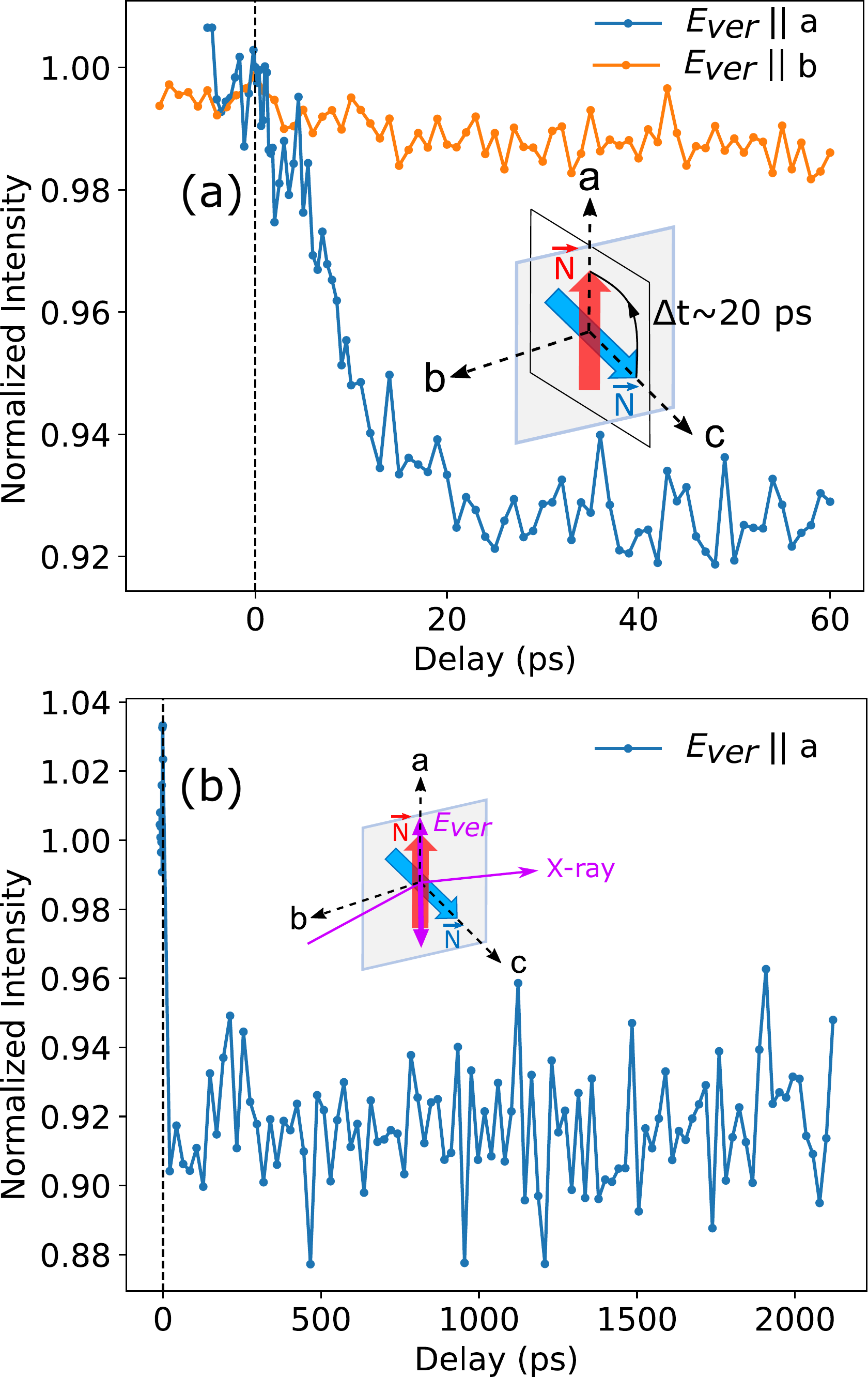}
\caption{(Colour Online) (a) Normalized pump-induced reflectivity change (pumped/unpumped) at $\sim$ 709\,eV as a function of pump–probe delay, measured with vertically polarized x-rays for two sample orientations: with the a-axis and b-axis aligned vertically. The inset illustrates the Néel vector trajectory, completing the spin reorientation within ~20\,ps. (b) Same measurement as in (a) for the vertical a-axis orientation, extended to longer pump–probe delays. The inset shows the sample geometry and x-ray polarization configuration. \label{Fig4}}
\end{figure}
To probe sub-picosecond dynamics of the SRT, we switched to slicing mode, set the photon energy to $\sim$ 709\,eV where the XMLD asymmetry is maximized, and varied the delay between the NIR pump and the X-ray probe to track the Néel vector dynamics. Note that in this mode, only vertical polarization is available.

We therefore aligned the sample such that the a-axis was oriented vertically (cf. inset of Fig. \ref{Fig4}(a)) and maintained the sample temperature well below 80 K ($\sim$~32 K) to ensure that the Néel vector remained along the c-axis in the ground state. 
If the Néel vector deviates from the c-axis, a corresponding change in intensity is expected.  
As shown in Fig. \ref{Fig4}(a) (blue curve), the normalized intensity (pumped/unpumped) decreases immediately upon pump arrival and continues to change until about 20 ps, after which it reaches a plateau that persists up to a pump-probe delay of 60 ps. 
Unlike in Fig. \ref{Fig3}(a), the pumped signal in this scenario decreases rather than increases as the initial orientation is reversed: here, the electric field ($E_\mathrm{ver}$) is perpendicular to the Néel vector in the ground state, whereas in Fig. \ref{Fig3}(a), $E_\mathrm{hor}$ is nearly parallel. Consequently, when the Néel vector transitions from the c-axis toward the a-axis, the signal change exhibits an opposite sign.

To determine the path of Néel vector switching, we reoriented the sample so that the b-axis is aligned vertically (cf. inset of Fig. \ref{Fig3}(b)) and measured the signal for the same pump-probe delays. If the Néel vector oscillated out of the a-c plane during reorientation, a corresponding XMLD asymmetry would be expected in this configuration within the first 20 ps due to the change of projection of the the Néel vector to the x-ray polarization direction. However, as shown in Fig. \ref{Fig4}(a) (orange curve), only a gradual decrease of signal of $\sim$ 1\% over 60 ps is observed, indicating that the Néel vector primarily rotates within the a-c plane. 

The inset of Fig. \ref{Fig4}(a) provides a schematic of the Néel vector’s dynamical trajectory as it transitions from the c-axis to the a-axis.

In order to investigate the recovery process, we realigned the sample to the former geometry (vertical a-axis) and extended the measurement to cover the maximum accessible pump-probe delay. The data, presented in Fig. \ref{Fig4}(b) (with the inset illustrating the measurement geometry), show no recovery even after more than 2 ns. We attribute this to slow heat dissipation due to the insulating nature of the sample and the relatively small absorption coefficient of about \SI{700}{\per\centi\meter} \cite{usachev2005}. This leads to a small temperature gradient at the surface of the crystal, which is predominately probed with the soft x-ray radiation. This distinguishes our work from a previous study, which reported a smaller Néel vector rotation of \SI{30}{\deg} \citet{Kimel2004}. Here, a \SI{60}{\micro\m}-thick sample was measured in transmission, averaging over a depth-dependent temperature gradient and a correspondingly varying rotation angle.
Our results indicate almost a complete rotation of \SI{90}{\deg}, as evidenced by the comparison between the XMLD asymmetry in static (cf. Fig. \ref{Fig2}) and time-resolved measurements (cf. Fig. \ref{Fig3}). With an estimated penetration depth of our probing radiation at \SI{5}{\deg} grazing incidence on the order of \SI{10}{nm}, we conclude that this surface region of the crystal exhibits almost a complete SRT.

\section{Conclusion}
We have investigated the Néel vector dynamics using time-resolved XMLD at grazing incidence in reflection, providing an X-ray perspective on the SRT in TmFeO$_3$. Unlike optical probes, which are sensitive to charge dynamics and can be affected by multiple extrinsic factors, our measurements directly track the local Fe moment. This provides an quantitative, artifact-free view of Néel vector dynamics on an ultrafast timescale.

The transient XMLD signals confirm that the Néel vector of TmFeO$_3$ primarily rotates within the a-c crystal plane and that the full \SI{90}{\deg} rotation is complete within \SI{20}{ps}. 
The laser-driven spin-reorientation remains unchanged for over \SI{2}{ns}, indicating slow thermal dissipation due to the insulating nature of the sample.  

Our study demonstrates the ability to probe element-specific AFM dynamics at their intrinsic timescale in reflection geometry, enabling the exploration of a wide range of AFM systems, including technologically relevant thin films. 

\section{Acknowledgement}
We acknowledge financial support from the Deutsche Forschungsgemeinschaft (DFG, German Research Foundation) – Project-ID 328545488 – TRR 227, project A02 and A03. R.K. acknowledge support by the Swedish Research Council
(VR 2021-5395) and the Knut and Alice Wallenberg Foundation (KAW Lightmatter). D. A. acknowledges the support of
ERC Grant No. 101078206 (ASTRAL)


\end{document}